%% file: frequency_attack_ver3.tex
\documentclass[conference]{IEEEtran}
\usepackage[margin=0.7 in]{geometry}
\usepackage{graphicx,epsfig,amsmath,amssymb,bm}
\usepackage{amssymb,balance}
\usepackage{amsmath}
\usepackage{amsthm} 
\usepackage{amssymb}
\usepackage{algorithmicx}
\usepackage{algpseudocode}
\usepackage{url}
\input{macros}

\usepackage{bbm}
\usepackage{subcaption}
\usepackage{tabularx}
\usepackage{multirow}
\usepackage{verbatim}
\usepackage[ruled,linesnumbered]{algorithm2e}

\usepackage{color}
\usepackage{tikz}
\usepackage[acronym]{glossaries}
\usepackage[capitalise]{cleveref}

\newcommand{\beqa}{\begin{eqnarray}}
\newcommand{\eeqa}{\end{eqnarray}}

\ifCLASSINFOpdf

\else

\fi

\newglossary[slg]{symbolslist}{syi}{syg}{List of Symbols}

\loadglsentries[symbolslist]{symbols}
\begin{document}

\title{A Comparison of Data-Driven Techniques for Power Grid Parameter Estimation}
\IEEEoverridecommandlockouts 
\author{
\IEEEauthorblockN{Subhash Lakshminarayana\IEEEauthorrefmark{1}, Saurav Sthapit\IEEEauthorrefmark{2} and Carsten Maple\IEEEauthorrefmark{3}} \\
\IEEEauthorblockA{\IEEEauthorrefmark{1}
School of Engineering, University of Warwick, UK \\
\IEEEauthorrefmark{2}
Warwick Manufacturing Group, University of Warwick, UK \\ 
Emails: \IEEEauthorrefmark{1}subhash.lakshminarayana@warwick.ac.uk, \IEEEauthorrefmark{2}Saurav.Sthapit@warwick.ac.uk, \IEEEauthorrefmark{3} cm@warwick.ac.uk}
}



\maketitle

\begin{abstract}
Power grid parameter estimation involves the estimation of unknown parameters, such as inertia and damping coefficients, using observed dynamics.
In this work, we present a comparison of data-driven algorithms for the power grid parameter estimation problem. First, we propose a new algorithm to solve the parameter estimation problem based on the Sparse Identification of Nonlinear Dynamics (SINDy) approach, which uses linear regression to infer the parameters that best describe the observed data. We then compare its performance against two benchmark algorithms, namely, the unscented Kalman filter (UKF) approach and the physics-informed neural networks (PINN) approach. We perform extensive simulations on IEEE bus systems to examine the performance of the aforementioned algorithms. Our results show that the SINDy algorithm outperforms the PINN and UKF algorithms in being able to accurately estimate the power grid parameters over a wide range of system parameters (including high and low inertia systems). Moreover, it is  extremely efficient computationally and so takes significantly less time than the PINN algorithm, thus making it suitable for real-time parameter estimation.
\end{abstract}

{\IEEEkeywords Parameter estimation, Sparse Identification of Nonlinear Dynamics, Unscented Kalman Filter, physics-informed neural networks.}


\IEEEpeerreviewmaketitle

\section{Introduction}
Accurate knowledge of the power system model and the underlying parameters plays a key role in maintaining reliable power grid operations. Power system parameters may change over time due to system aging and varying operational conditions. Moreover, the growing integration of renewable energy resources, demand-response programs, storage, etc. introduces significant variation and uncertainty in system parameters (such as inertia). The lack of accurate power system models and parameter knowledge can lead to incorrect assumptions about system safety. Indeed, this has lead to real-world system blackout incidents such as the 1996 Western North American outage \cite{WSCC99}.
Real-time tracking of the system parameters is thus extremely important. 
Phasor measurement units (PMUs) provide fine-grained monitoring of power grid signals (such as voltage phase angle, frequencies, etc.). They can be leveraged to track the system parameters in an online manner. Given the growing deployment of PMUs, data-driven real-time parameter estimation algorithms are being increasingly proposed \cite{AriffUKF2015, Terzija2011, Abur2018, Huang2018, stiasny2020PINN}.

The power grid parameter estimation problem has, to date, been primarily addressed through the application of the Kalman filter approach and its variants \cite{ZhaoGomez2019}. The basic idea behind this approach is to perform joint state and parameter estimation, i.e., augment the unknown system parameters to the state variables for joint estimation. Several works follow this approach. In particular, \cite{AriffUKF2015} applied a UKF based approach to estimate the transient reactance and inertia of the generator. Extensions of the UKF approach were used to estimate the system parameters for higher-order generator models in \cite{Terzija2011} and \cite{Abur2018}.
Reference \cite{Huang2018} proposed a tool to calibrate models/parameters based on trajectory sensitivity analysis and an ensemble-KF approach. However, KF-based techniques cannot perform well in the presence of strong non-linearity and for low-inertia systems that exhibit very fast dynamics \cite{stiasny2020PINN}.

To overcome the drawbacks of the KF-based approach, reference \cite{stiasny2020PINN} presented a data-driven approach for power grid parameter estimation based on the technique of physics-informed neural networks (PINNs). The PINN algorithm uses the knowledge of the physical model and the observed power grid dynamical signals to train a neural network (NN) in real-time to infer the state and unknown parameters of the system. Since it is based on NNs, it is capable of handling strong non-linearity. However, despite these favourable properties, the PINN method has drawbacks. First, training a NN is computationally complex, and will be time-consuming, especially for large power grids. Hence, it may not be suitable for real-time parameter estimation. Moreover, as pointed in \cite{stiasny2020PINN}, the PINN algorithm is not suitable when estimating parameters in high-inertia systems that exhibit very slow dynamics, due to difficulties in training the NN.

To overcome these limitations, in this work, we propose a new algorithm for data-driven power grid parameter estimation based on the Sparse Identification of Nonlinear Dynamics (SINDy) approach \cite{BruntonSINDy2016}, which applies linear regression to determine the power grid parameters that best describe the observed measurements. The main idea behind this algorithm applied in our context can be summarized succinctly as follows. Consider a system described by a simple dynamical model of the form $\dot{x} = a x,$ where $x$ is the state of the system and $a$ is an unknown parameter to be estimated. Assume that the system operator has a sensor deployed that monitors a sampled and noisy version of the state variable $x^{(\tau)}, \tau = 1,\dots,T,$ where $T$ is the time horizon of observation. The algorithm proceeds by computing the numerical derivative of the signal $x$ to obtain the values of $\dot{x}$ from $x.$ If the data is noiseless, the numerical derivatives can be computed using the finite differences method. However, for noisy data, the polynomial interpolation is better-suited \cite{BruntonSINDy2016}. Then, based on the knowledge of time series data of $x^{(\tau)}$ and $\dot{x}^{(\tau)},$ the estimation of the unknown parameter $a$ can be set up as a linear regression problem. We note that although the SINDy algorithm described above is based on linear regression, the method is broadly applicable for non-linear systems \cite{BruntonSINDy2016} (we show this subsequently for the case of a non-linear power grid model). If the system operator does not have knowledge of the exact differential equation describing the physical system, then the SINDy algorithm can still be applied by constructing a library of candidate linear/non-linear functions of the observed state and performing a sparse regression to select the terms that best describe the observed dynamics. 

The key contributions of this work are as follows.
\begin{itemize}
    \item We develop the SINDy approach for estimating the inertia and damping coefficients for a non-linear power grid model.
    \item We show that the proposed SINDy algorithm can be implemented in a decentralized manner that only requires exchanging the observed phase angle data between the neighbouring nodes.
    \item We conduct extensive simulations using IEEE bus systems to evaluate and compare the performance of these algorithms.
\end{itemize}
Our results show that the SINDy algorithm performs well over a wide range of power grid system parameters, i.e., for low and high inertia systems that exhibit fast as well as slow dynamics. This is in contrast to the PINN algorithm that has a large estimation error for systems with slow dynamics and the UKF algorithm that has a large error for systems with fast dynamics.  Moreover, since the SINDy algorithm is based on computationally simple linear regression, the computation time is extremely fast. Thus, it is suitable for implementation for real-time parameter estimation in large power systems.

The rest of the paper is organized as follows. Section~\ref{sec:Prelim} introduces the system model and the parameter estimation problem. 
Section~\ref{sec:Algos} describes the SINDy algorithm for power grid parameter estimation and briefly reviews the PINN method. Section~\ref{sec:Sims} describes the simulation results and Section~\ref{sec:Conc} concludes.

\section{Preliminaries}
\label{sec:Prelim}
\subsection{System Model}
We consider a power grid consisting of a set of $\mathcal{N} = \{1,\dots,N \}$ buses. The buses are divided into generator buses $\mathcal{N}_G$ and load buses $\mathcal{N}_L$ and $\mathcal{N} = \mathcal{N}_G \cup \mathcal{N}_L$. The power grid dynamic model is given by  \cite{kundur1994}:
\begin{align}
  \dot{\delta}_i & = \omega_i,  i \in \mathcal{N}_G \label{eqn:dyn1} \\
 M_i \dot{\omega}_i & = - D_i \omega_i + P^M_i   - \sum_{j \in \mathcal{N}} B_{i,j} \sin (\delta_{ij}),  \ i \in \mathcal{N}_G, \label{eqn:dyn2}\\
D_i \dot{\delta}_i &=   - P^{L}_i  - \sum_{j \in \mathcal{N}} B_{i,j} \sin (\delta_{ij}), \ i \in \mathcal{N}_L, \label{eqn:dyn3}
\end{align}
where $\delta_i$ is the phase angle deviation at bus $i \in \mathcal{N},$ 
$\delta_{ij} = \delta_i - \delta_j, \ i,j \in \mathcal{N}$ and
$\omega_i$ denotes the rotor frequency deviation at the generator buses $i\in \mathcal{N}_G.$ The generator inertia coefficient at bus $i \in \mathcal{N}_G$ is denoted by $M_i$ and the mechanical input power by $P^M_i.$ The damping coefficients at generator/load buses is denoted by 
$D_i$ $i \in \mathcal{N}$. 
$B_{i,j}$ is the susceptance of line $i,j.$ Finally, $P^L_i $ denotes the load at bus $i \in \mathcal{N}_L$. 




\subsection{Power Grid Parameter Estimation Problem}
In this work, we focus on estimating the inertia and damping coefficients, $M_i, i \in \mathcal{N}_G$ and $D_i, i \in \mathcal{N}$ by observing the power grid dynamics. We assume that the system operator has deployed PMUs in the grid, which enables them to monitor the voltage phase angles $\{ \delta^{(\tau)}_i \}_{i \in \mathcal{N}, m = 1,\dots,T}$ and frequency fluctuations $\{ \dot{\delta}^{(\tau)}_i \}_{i \in \mathcal{N}, \tau = 1,\dots,T}$ respectively over a period of time. Herein, we assume a slotted time system with $x^{(\tau)}$ denoting the value of the signal $x$ at time slot $\tau,$ where the slots are sampled at a time interval of $T_s,$ and $T$ is the total number of time slots. For instance, according to IEEE/IEC standards, for a $50~$Hz system, the PMU sampling frequency can be between 10 to 100 frames per second. Therefore, $T_s$ is in the range $10-100~$ms \cite{PMU2018}.

\section{Data-Driven Techniques for Power Grid Parameter Estimation}
\label{sec:Algos}
The focus of this work is on data-driven algorithms to solve the power grid parameter estimation problem. We note that the aforementioned problem can be treated as a supervised learning problem in which we train a machine learning classifier offline under various system parameter settings. The trained model can be used to perform online inference of the underlying parameters. However, such a \emph{black-box} approach would require a combinatorially large search across various power grid parameters that must be estimated, and hence, computationally complex.

Alternately, in this section, we propose a novel data-driven algorithm to solve the parameter estimation problem based on the SINDy algorithm \cite{BruntonSINDy2016}, which leverages on the knowledge of the physical model under consideration (i.e., the differential equations describing the physical system) to identify the system parameters. We also briefly describe the PINN approach to estimate the system parameters \cite{stiasny2020PINN}, which also similarly uses the knowledge of the system model.

\subsection{Parameter Estimation Using the SINDy Algorithm}
The SINDy algorithm applies linear regression to find the system parameters that best represent the observed power grid dynamics \cite{BruntonSINDy2016}. In order to formalize this framework for power grid parameter estimation, we define the following additional notations. Let $a^{(\tau)}_i $ and $b^{(\tau)}_i$ be given by
\begin{align}
a^{(\tau)}_i & =   P^{M}_i + \sum_{j \in \mathcal{N}} B_{i,j} \sin (\delta^{(\tau)}_{ij}),  \label{eqn:rearr1} \\
b^{(\tau)}_i & = P^{L}_i + \sum_{j \in \mathcal{N}} B_{i,j}  \sin (\delta^{(\tau)}_{ij}). \label{eqn:rearr2}
\end{align}
Using \eqref{eqn:rearr1} and \eqref{eqn:rearr2}, a discretized version of the equations describing the generator and the load bus dynamics in \eqref{eqn:dyn2} and \eqref{eqn:dyn3}  over $\tau = 1,2,\dots,T$ time slots can be written as 
\begin{align}
\underbrace{\begin{bmatrix}
\dot{\omega}^{(1)}_i \\
    \vdots  \\
\dot{\omega}^{(T)}_i \\ 
\end{bmatrix}}_{\dot{\omegav}_i}  &= \underbrace{ \begin{bmatrix}
-\omega^{(1)}_{i},-a^{(1)}_i \\
    \vdots  \\
-\omega^{(T)}_{i},-a^{(T)}_i \\ 
\end{bmatrix}}_{\Omega } \underbrace{\begin{bmatrix}
\frac{D_i}{M_i} \vspace{0.3 cm} \\
\frac{1}{M_i}
\end{bmatrix}}_{\pv^G}, i \in \mathcal{N}_G, \label{eqn:Regression_full_gen} \\
\underbrace{\begin{bmatrix}
\dot{\delta}^{(1)}_i \\
    \vdots  \\
\dot{\delta}^{(T)}_i \\ 
\end{bmatrix}}_{\dot{\deltav}_i}  &= \underbrace{ \begin{bmatrix}
-b^{(1)}_i \\
    \vdots  \\
-b^{(T)}_i \\ 
\end{bmatrix}}_{\bv} \underbrace{\begin{bmatrix}
\frac{1}{D_i}
\end{bmatrix}}_{\pv^L}, i \in \mathcal{N}_L. \label{eqn:Regression_full}
\end{align}
Equations \eqref{eqn:Regression_full_gen} and \eqref{eqn:Regression_full} represent the relationship between the generator/load inertia, damping coefficients (that must be estimated) and the observed PMU data. 
We represent the matrices on the right-hand side of \eqref{eqn:Regression_full_gen} and \eqref{eqn:Regression_full} by $\Omega$ and $\bv$ respectively, and the generator and load parameters by $\pv^G$ and $\pv^L$ respectively.
Note that the system operator can compute the elements of $\Omega$ and $\bv$ matrices using the frequency and phase angle measurements obtained from the PMUs. 
Using these measurements, the left-hand side of \eqref{eqn:Regression_full_gen} and \eqref{eqn:Regression_full} can be computed by taking numerical derivatives of $\dot{\delta}^{(\tau)}_i, \ i \in \mathcal{N}_G$ and $\delta^{(\tau)}_i, \ i \in \mathcal{N}_L$. For noiseless data, the derivative can be computed using the finite difference method. For noisy data, the polynomial interpolation method is better suited \cite{BruntonSINDy2016}.

 Equations \eqref{eqn:Regression_full_gen} and \eqref{eqn:Regression_full} are a system of $T$ linear equations. The unknown parameters $\pv^G$ and $\pv^L$ can be recovered by solving the linear regression problem. Further, the operator can recover the values of $M_i$ and $D_i$ from the estimated value of $\pv^G$ in a straightforward manner. We note that although the framework presented above involves linear regression, it can be applied to power grid system identification problem involving non-linear system model, such as the one presented in \eqref{eqn:dyn1}-\eqref{eqn:dyn3}, since the terms inside the non-linearity can be measured (e.g., the operator can compute $\sin(\delta^{(\tau)}_{i,j})$ using the measurements $\delta^{(\tau)}_{i,j}.)$ Thus, the SINDy algorithm is applicable for parameter estimation in general non-linear power system models.

Finally, in the algorithm described above, we assume that the system operator has the precise knowledge of the differential equations governing the power grid dynamics (i.e., equations \eqref{eqn:dyn1}-\eqref{eqn:dyn3}). However, this knowledge may not be available precisely for many real-world systems. In such cases, the system operator can create a library of candidate linear and non-linear terms based on the observed data to construct the right-hand side matrices of \eqref{eqn:Regression_full_gen} and \eqref{eqn:Regression_full} respectively (i.e., $\Omega$ and $\bv$). For instance, to estimate the generator parameters, the operator can create a large library of functions such as
\begin{align}
    \Um = [ \fv_1(\omegav^{(\tau)},\av^{(\tau)}), \fv_2(\omegav^{(\tau)},\av^{(\tau)}),\dots,],
\end{align}
where $\fv_1(.),\fv_2(.),\dots,$ are arbitrary candidate linear/non-linear functions, such as $\omegav^{(\tau)}$, $(\omegav^{(\tau)})^2$, $\sin(\av^{(\tau)}), \cos(\av^{(\tau)}), \dots.$ Then, the generator parameters $\pv^G$ can be estimated as a sparsest solution that satisfies the following equation
\begin{align}
    \dot{\omegav}_i = \Um \pv^G.
\end{align}
In this case, the sparse solution determines which of the candidate terms are contributing to the observed dynamics. Moreover, the sparse solution also avoids an NP-hard combinatorial brute-force search across all possible term combinations.

\subsection{Decentralized Implementation}
The proposed SINDy algorithm is suitable for decentralized implementation. 
Recall that the algorithm involves solving for the unknown terms in \eqref{eqn:Regression_full_gen} and \eqref{eqn:Regression_full}. This in turn requires the following signals at each node $i \in \mathcal{N}$:  (i) $\omega_i,$ (ii) $\delta_i,$ (iii) $ \{ \delta_{j} \}_{j \in \mathcal{N}_i},$ where $\mathcal{N}_i$ are the neighbouring nodes of node $i.$ Note that (i) and (ii) can be monitored locally at individual nodes, and (iii) only requires information exchange with only the neighboring nodes. Thus, the SINDy algorithm can be implemented locally with limited information exchange between the nodes.

\subsection{Parameter Estimation Based on Physics-Informed Neural Networks}
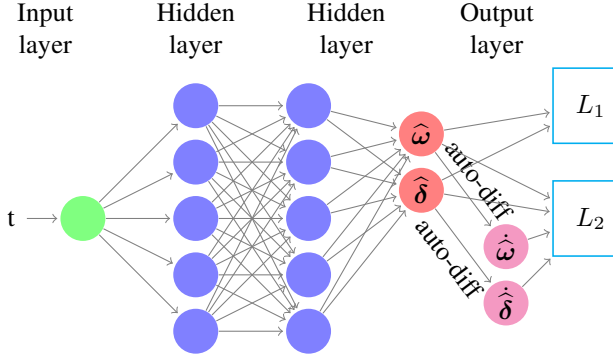
\begin{figure}
    \centering
    \input{fig_pinn}
    \caption{PINN network showing input output and the losses}
    \label{fig:pinn_arch}
\end{figure}

We also briefly describe the  PINN framework  \cite{RaissiPINN2018, Lagaris1998} applied to power system parameter estimation as proposed in \cite{stiasny2020PINN}. 
The overall PINN framework is shown in Fig.~\ref{fig:pinn_arch}. It uses NNs to approximate the solution ${\omega_i} (t)$ and ${\delta_i} (t)$
that solves \eqref{eqn:dyn1}-\eqref{eqn:dyn3}. Let 
$\widehat{\omega}_i (t,\phiv)$ and $\widehat{\delta}_i (t,\phiv)$ denote the 
respective NN approximations as shown in Fig.~\ref{fig:pinn_arch}. Herein, $\phiv$ represents the weights of the NN, which are trained to minimize the following losses: 

\subsubsection{Mean squared loss}
The first loss term involves minimizing the mean square loss between the observed measurements ${\omega_i} (t)$ and ${\delta_i} (t)$ and their NN approximations, i.e., 
\begin{align}
    L_1 (\phiv) =\frac{1}{T}\sum_{\tau=1}^{T} & \Big{(} \sum_{i \in \mathcal{N}_G} (\widehat{\omega}_i(\tau ,\phiv) - {\omega}_i(\tau))^2 \nonumber  \\ & + \sum_{i \in \mathcal{N}} (\widehat{\delta}_i(\tau,\phiv) - {\delta}_i(\tau))^2 \Big{)}.
\end{align}
\subsubsection{Physics based Loss}
The system dynamics are incorporated as a second loss function. Specifically, let us define
\begin{align}
 f^{(1)}_i & =  \dot{\widehat{\delta}}_i  - \widehat{\omega}_i,  i \in \mathcal{N}_G \label{eqn:dyn11} \\
f^{(2)}_i & =  M_i \dot{\widehat{\omega}}_i   + D_i \widehat{\omega}_i - P^M_i 
 \nonumber \\ & \qquad \qquad 
 + \sum_{j \in \mathcal{N}} B_{i,j} \sin (\widehat{\delta}_{ij}),  \ i \in \mathcal{N}_G, \label{eqn:dyn12}\\
f^{(3)}_i & = D_i \dot{\widehat{\delta}}_i + P^{L}_i  + \sum_{j \in \mathcal{N}} B_{i,j} \sin (\widehat{\delta}_{ij}), \ i \in \mathcal{N}_L, \label{eqn:dyn13}
\end{align}
where in the above, we have dropped the notations showing the dependency of $f^{(1)}_i$ and $f^{(2)}_i$ on $(\tau,\phiv)$ and $f^{(3)}_i$ on $(\tau, \phiv, \pv),$   where $\pv = [\pv^G;\pv^L],$ and simplicity. Note that in the above, the derivatives are computed using NN's automatic differentiation. The physics-based loss function is then given by
\begin{align}
    L_2  & (\phiv, \pv)   =\frac{1}{T}\sum_{\tau=1}^{T} \Big{(} \sum_{i \in \mathcal{N}_G} (f^{(1)}_i (\tau, \phiv))^2  \nonumber \\ &  +  \sum_{i \in \mathcal{N}_G} (f^{(2)}_i(\tau, \phiv)^2 +  \sum_{i \in \mathcal{N}_L} (f^{(3)}_i(\tau, \phiv, \pv)^2 \Big{)}.
\end{align}
Note that the above loss function depends both on the NN weights $\phiv$ and the system parameters $\pv.$
 The NN is trained using the observed measurements  $\{ \delta^{(\tau)}_i \}_{i \in \mathcal{N}, \tau = 1,\dots,T}$ and $\{ \omega^{(\tau)}_i \}_{i \in \mathcal{N}, \tau = 1,\dots,T}$ as follows:
\begin{align}
  \phiv^*, \pv^* = \arg \min_{\phiv, \pv } L_1  (\phiv) + L_2  (\phiv, \pv)
\end{align}
Note that when the sum of these two losses is minimized, 
we ensure two criteria: (1) the output of the NN replicates the observed system dynamics, (2)  $f^{(1)}_i, f^{(2)}_i,$ and $f^{(3)}_i$ are close to zero, which in turn implies that \eqref{eqn:dyn1}-\eqref{eqn:dyn3} are satisfied. This implies that the estimated parameters best fit the observed data.

\section{Simulation Results}
\label{sec:Sims}
In this section, we present simulation results to show the effectiveness of the SINDy algorithm and compare its performance against PINN and UKF algorithms.

For our simulations, we obtain the power grid topological data from the MATPOWER simulator. To generate the dataset corresponding to power grid dynamics, we simulate the differential equations in \eqref{eqn:dyn1} - \eqref{eqn:dyn3} using the \emph{ode-45} function of MATLAB. We use the polynomial interpolation method to take numerical derivatives and solve the linear regression problems using MATLAB. 
For the PINN algorithm, we use the code provided in \cite{stiasny2020PINNcode} that is implemented using Tensorflow. We use a multi-layer perceptron for PINN with two hidden layers each with $30$ neurons and the automatic differentiation to calculate the gradients. The NN is trained using ADAM optimizer.

\begin{figure}[!t]
\centering
\includegraphics[width=0.48\textwidth]{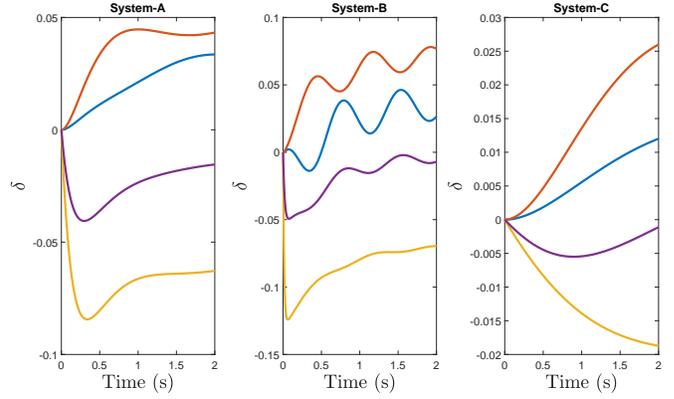}
\caption{Frequency dynamics for the 4-bus system.}
\label{fig:freq_dyn_plot39bus}
\vspace{-0.2 cm}
\end{figure}

We first compare the performance of the algorithms using the 4-bus 2-generator model. We use three simulation settings as in \cite{stiasny2020PINN} -- system A that is considered as the standard system, system B with faster dynamics (as compared to system A), and system C with slower dynamics. The parameter values are listed in the Appendix. The phase angle dynamics corresponding to the three settings are shown in \cref{fig:freq_dyn_plot39bus}. To generate a noisy signal, we add Gaussian noise whose standard deviation $\sigma$ is set to $5 \%$ of the measurement's value. We consider $100$ samples per second (thus, a PMU sampling interval of $0.01$ seconds) and a time horizon of $2$ seconds. Thus, $ T = 2 \times 100 = 200.$

\begin{figure}[!t]
\centering
\begin{subfigure}{0.45\textwidth}
\includegraphics[width=0.97\columnwidth]{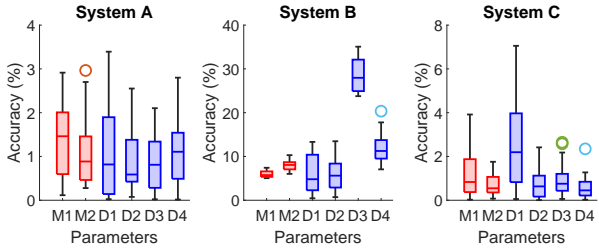}
\caption{SINDy algorithm.}
\vspace{0.5 cm}
\end{subfigure}
~
\begin{subfigure}{0.45\textwidth}
\includegraphics[width=0.97\columnwidth]{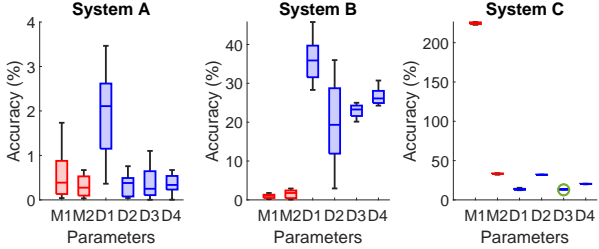}
\caption{PINN algorithm.}
\end{subfigure}
\caption{Error in parameter estimation for 4-bus 2-generator model. Red and blue boxes represent the error in unknown parameters $M$ and $D$ respectively.}
\label{fig:accuracy_SR_4bus}
\end{figure}

The error in parameter estimation (boxplots) considering the 4-bus 2-generator model 
is shown in Fig.~\ref{fig:accuracy_SR_4bus} for the  SINDy and PINN algorithms respectively over $20$ simulation runs. As evident from the figure, both the PINN and the SINDy algorithms have a high estimation accuracy for System A, and most of the unknown parameters are estimated within $3-4$ percent of the original value. 
Both the algorithms also have similar performance for System B, i.e., the system fast dynamics. We observe that $M$ parameters are recovered with good accuracy but the $D$ parameters have higher error (within $40\%$). However, in terms of the actual value, the true and estimated parameters are very close, e.g., $D_4 = 0.02$ and $\hat{D}_4 = 0.028.$ Thus, we do not expect this error to significantly affect power grid operations.

The most notable difference between the two algorithms can be seen for System~C that has slow dynamics.
While the PINN algorithm has a very high estimation error for some parameters (up to $200 \%$), the SINDy algorithm has a very low estimation error for this case as well. As noted in \cite{stiasny2020PINN}, the large error for the PINN algorithm can be attributed to the challenges in training the NN. Specifically, for the system with slow dynamics, the optimization landscape is very flat, and hence, no direction of the gradient dominates. In contrast, the SINDy algorithm, which is based on a computationally simple linear regression faces no such obstacles.

For a comparison with the UKF approach, we refer to the results in \cite{stiasny2020PINN}, where it is noted that the UKF algorithm does not perform well for System~B (with fast dynamics). This is because the UKF algorithm cannot capture fast information dynamics that occur within a few time steps. Thus, we conclude that the 
SINDy algorithm has a better performance compared to the UKF and the PINN approach over a wide range of system parameters.

Another significant advantage of the SINDy algorithm is its fast computing time. The simulations for the SINDy algorithm are conducted using a Macbook Pro 2.3 GHz Dual-Core Intel Core i5 processor with 8GB RAM and we enlist the run time in Table~\ref{tbl:Exe_Time}. It can be noted that the SINDy algorithm is computationally extremely fast compared to the PINN algorithm (by about $1000$ times for the 4-bus system) since it does not involve training a neural network. Furthermore, from the table, we note that the computing time is also extremely low for larger bus systems (results to be enlisted next). Thus, this approach is better suited for real-time parameter estimation.

\begin{table}[!t]
 \begin{center}
 \begin{tabular}{|c | c | c | c | } 
 \hline
  & Execution time (s)  \\ [0.5ex] 
 \hline\hline
PINN (4 bus)  & $90$ s (from \cite{stiasny2020PINN})  \\ \hline
SINDy (4 bus) & $8.4~$ms   \\ \hline
SINDy (IEEE 6-bus) & $10.5~$ms  \\ \hline
SINDy (IEEE 39-bus) & $86~$ms  \\
 \hline
\end{tabular}
\end{center}
\caption{Computation times for the SINDy and PINN algorithms.}
 \label{tbl:Exe_Time}
\end{table} 
\subsubsection*{Dependence on the Observation Time Window}
Next, we examine the dependency of the estimation error on the observation time window for the SINDy algorithm. Once again, we consider the 4-bus 2-generator system. In Table~\ref{tbl:obs_Time}, we enlist the mean accuracy of $D_1$ over $20$ trials (we observed that the estimation error for this parameter is the greatest among all the parameters). We also consider two different noise levels, i.e., $\sigma = [5 \%, 10 \%]$ of the measurement's value. It can be noted that the algorithm is able to achieve low estimation error with data from about $2$ seconds observation time window. Furthermore, more noisy data requires a slightly larger observation time window, and the estimation error is slightly higher, which is expected.

\begin{table}[!t]
 \begin{center}
 \begin{tabular}{|c | c | c | c | c| } 
 \hline
 Time window (s) & $\sigma = 5 \%$ & $\sigma = 10 \%$ \\ [0.5ex] 
 \hline\hline
$0.5$  & $22.2 \%$  & $40.3 \%$ \\ \hline
$1$ & $12.2 \%$ & $21.84 \%$ \\ \hline
$2$ sec & $1.12 \%$  & $2.82 \%$ \\ \hline
$5$ sec & $1 \%$ & $2.46 \%$ \\
 \hline
\end{tabular}
\end{center}
\caption{Mean estimation error of $D_1$ as a function of the observation time window and noise level for the SINDy algorithm over $20$ runs. The results correspond to the 4-bus 2-generator system, System~A.}
 \label{tbl:obs_Time}
\end{table} 

\subsubsection*{Simulations with Larger Bus Systems}
\begin{figure}
    \centering
    \includegraphics[width=0.7\columnwidth]{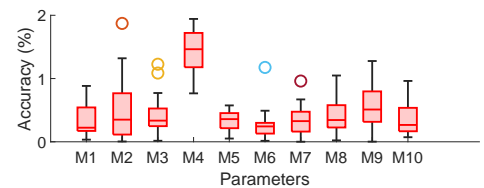}
    \caption{Error in parameter estimation for IEEE-39 bus system. }
    \label{fig:accuracy_39bus}
\end{figure}

We also plot the error in parameter estimation for the IEEE-6 bus and 39 bus systems. For brevity, we only plot the accuracy of the inertia estimations ($M_i$) for the IEEE-39 bus system (note the 39-bus system has $10$ generators and $29$ loads). The simulation parameters are once again listed in the Appendix and the results are plotted in Fig.~\ref{fig:accuracy_6bus} and Fig.~\ref{fig:accuracy_39bus}. It can be observed that the SINDy approach achieves high accuracy even for larger bus systems. On the other hand, we did not observe good estimation results for the PINN algorithm (as shown in Fig.~\ref{fig:accuracy_6bus}~(b)) using the code provided in \cite{stiasny2020PINNcode}. We believe further data processing may be required to ensure the scalability of the PINN algorithm to large systems.

\begin{figure}[!t]
\centering
\begin{subfigure}{0.45\textwidth}
\centering
\includegraphics[width=0.77\columnwidth]{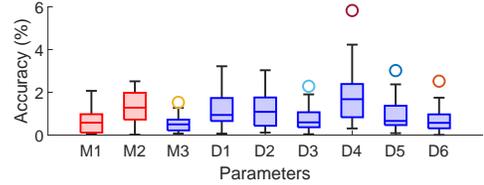}
\caption{SINDy algorithm.}
\vspace{0.5 cm}
\end{subfigure}
~
\begin{subfigure}{0.45\textwidth}
\centering
\includegraphics[width=0.77\columnwidth]{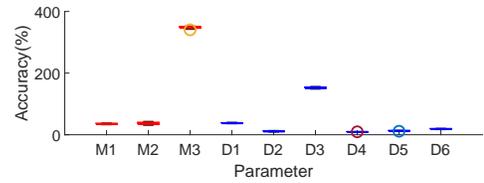}
\caption{PINN algorithm.}
\end{subfigure}
\caption{Error in parameter estimation for IEEE-6 bus system. Red and blue boxes represent the error in unknown parameters $M$ and $D$ respectively.}
\label{fig:accuracy_6bus}
\end{figure}

\subsubsection*{Simulations with Non-Gaussian Noise}
We also conducted simulations considering Non-Gaussian noise, since some recent work based on real-world experiments shows that PMU noise data can be better modeled Student-t or Logistic distributions \cite{Wang2018}.
We added Logistic noise whose parameters are adjusted to obtain a standard deviation of $0.01$ pu. We observed that the performance of the SINDy algorithm under non-Gaussian noise conditions is very similar to the Gaussian case. For brevity, exclude the results here, since they are similar to Fig.~\ref{fig:accuracy_SR_4bus}.

\section{Conclusions}
\label{sec:Conc}
In this work, we propose a novel data-driven algorithm to estimate power grid parameters from the observed phase angle/frequency data. The proposed method, based on the SINDy algorithm, applies linear regression to estimate the power grid parameters that best describe the observed data. The algorithm is applicable for general non-linear power grid models. Simulation results conducted on standard power grid bus systems show that the proposed algorithm performs well under various parameter settings, and bus systems. Our future work includes an analysis of the optimal placement of PMUs to ensure reliable parameter estimation and application to higher-order generator models.

\balance
\bibliographystyle{IEEEtran}
\bibliography{IEEEabrv,bibliography}

\section*{Appendix}
All the values are provided in pu.
\subsection{4-bus system} 
$P^M_1 = 0.1, P^M_2 = 0.2,$ $P^L_1 = 0.1, P^L_2 = 0.2$

\vspace{0.1 cm}
{\emph System A:} \\
$M_1=0.3,M_2=0.2$\\
$D_1-D_4:0.15,0.3,0.25,0.25$\\

{\emph System B:} \\
$M_1=0.02,M_2=0.03$\\
$D_1-D_4:0.015,0.015,0.02,0.04$\\

{\emph System C:} \\
$M_1=5.2,M_2=4.0$\\
$D_1-D_4:3.8,4.3,10.5,8.3$\\

\subsection{IEEE 6-bus system} 
\noindent $M_1-M_3: 1.25,0.34,0.16$ \\
$D_1-D_6: 1.25,0.68,0.32,1,1,1$ \\
$P^M_1 = 0.2, P^M_2 = 0.1,$ $P^L_1 = 0.1, P^L_2 = 0.2$

\subsection{IEEE 39-bus system} 
 \noindent $M_{1} = 2.3186,   M_{2} : M_{8} = 2.6419,  M_{9} : M_{10}  = 2.4862.$ \\
 $D_i = 2,  \forall i \in \mathcal{N}_G$ \\
$D_i = 0.1, \forall i \in \mathcal{N}_L$ \\
$P^M_{33} = 0.2,$ $P^L_{19} = 0.2.$

\end{document}

%% file: macros.tex
\newfont{\bbb}{msbm10 scaled 500}

\newfont{\bb}{msbm10 scaled 1100}

\newcommand{\av}{{\bf a}}
\newcommand{\bv}{{\bf b}}

\newcommand{\fv}{{\bf f}}

\newcommand{\pv}{{\bf p}}

\newcommand{\Um}{{\bf U}}

\newcommand{\deltav}{\hbox{\boldmath$\delta$}}

\newcommand{\phiv}{\hbox{\boldmath$\phi$}}

\newcommand{\omegav}{\hbox{\boldmath$\omega$}}

\renewcommand{\arg}{{\hbox{arg}}}

%% file: symbols.tex
\newglossaryentry{unknown_param}{name={\ensuremath{\lambda}},description={Unknown attack parameter of $K^L$ matrix}}
\newglossaryentry{regulariser_penalty}{name={\ensuremath{\alpha}},description={Penalty factor for regularisation}}

%% file: fig_pinn.tex
\def\layersep{2cm}

\begin{tikzpicture}[shorten >=1pt,->,draw=black!50, scale=0.75,node distance=\layersep]
    \tikzstyle{every pin edge}=[<-,shorten <=1pt]
    \tikzstyle{neuron}=[circle,fill=black!25,minimum size=17pt,inner sep=0pt]
    \tikzstyle{input neuron}=[neuron, fill=green!50];
    \tikzstyle{output neuron}=[neuron, fill=red!50];
    \tikzstyle{deriv neuron}=[neuron, fill=magenta!50];
    \tikzstyle{hidden neuron}=[neuron, fill=blue!50];
    \tikzstyle{annot} = [text width=4em, text centered]

    \foreach \name / \y in {1}
        \node[input neuron, pin=left:t] (I-\name) at (0,-2.5) {};

    \foreach \name / \y in {1,...,5}
        \path[yshift=0.5cm]
            node[hidden neuron] (H-\name) at (\layersep,-\y cm) {};
    \foreach \name / \y in {1,...,5}
        \path[yshift=0.5cm]
            node[hidden neuron] (H1-\name) at (4cm,-\y cm) {};
        \path[yshift=0.5cm]
            node[output neuron] (Oomega) at (6cm,-1.5 cm) {$\widehat{\omegav}$};
        \path[yshift=0.5cm]
            node[output neuron] (Odelta) at (6cm,-2.5 cm) {$\widehat{\deltav}$};
            \path[yshift=0.5cm]
    
            node[deriv neuron] (Oomegadot) at (7.5cm,-3.5 cm) {$\dot{\widehat{\omegav}}$};
            \path[yshift=0.5cm]
            node[deriv neuron] (Odeltadot) at (7.5cm,-4.5 cm) {$\dot{\widehat{\deltav}}$};
    \node[ semithick,draw=cyan,minimum size=1cm] (l1) at (9cm,-0.5cm) {$L_1$};
    \node[ semithick,draw=cyan,minimum size=1cm] (l2) at (9cm,-2.5cm) {$L_2$};

    \draw[->] (Oomega) -- (Oomegadot) node [midway, above, sloped] (TextNode1) {auto-diff};
    \draw[->] (Odelta) -- (Odeltadot)  node [midway, below, sloped] (TextNode2) {auto-diff};
    \path (Oomega) edge (l1);
    \path (Odelta) edge (l1);
    \path (Oomega) edge (l2);
    \path (Odelta) edge (l2);
    \path (Oomegadot) edge (l2);
    \path (Odeltadot) edge (l2);
    
    \foreach \source in {1}
        \foreach \dest in {1,...,5}
            \path (I-\source) edge (H-\dest);
    \foreach \source in {1,...,5}
        \foreach \dest in {1,...,5}
            \path (H-\source) edge (H1-\dest);
    \foreach \source in {1,...,5}
        \foreach \dest in {omega,delta}
        \path (H1-\source) edge (O\dest);

    \node[annot,above of=H-1, node distance=1cm] (hl) {Hidden layer};
    \node[annot,left of=hl] {Input layer};
    \node[annot,right of=hl](h1l) {Hidden layer};
    \node[annot,right of=h1l] {Output layer};
\end{tikzpicture}